\documentclass[%
 reprint,
 amsmath,amssymb,
 aps,
 pra,
floatfix,
]{revtex4-2}

\usepackage{graphicx}
\usepackage{dcolumn}
\usepackage{bm}
\usepackage[normalem]{ulem}

\usepackage[
range-units = single,
]{siunitx}
\usepackage{xspace}
\usepackage{xcolor}
\usepackage[colorlinks=true, allcolors=blue]{hyperref}

\usepackage{xstring}
\newcommand*{\aref}[1]{%
	\IfBeginWith{#1}{eq:}{Eq.~\eqref{#1}}{}%
	\IfBeginWith{#1}{fig:}{Fig.~\ref{#1}}{}%
	\IfBeginWith{#1}{tab:}{Table~\ref{#1}}{}%
	\IfBeginWith{#1}{appendix:}{Appendix~\ref{#1}}{}%
	\IfBeginWith{#1}{sec:}{Section~\ref{#1}}{}%
}

\usepackage{subfiles}

\newcommand{\be}{\begin{align}}
\newcommand{\ee}{\end{align}}
\newcommand{\ket}[1]{\ensuremath{\left| {#1} \right>}}

\newcommand{\nbar}{\bar{n}}

\newcommand{\bC}[1]{\textcolor{black}{#1}}

\newcommand{\Be}{\ensuremath{^9\mathrm{Be}^+\;}}
\newcommand{\BeNoSpace}{\ensuremath{^9\mathrm{Be}^+}}

\newcommand{\Htwop}{\ensuremath{\mathrm{H}_2^+\;}}

\newcommand{\HtwopNoSpace}{\ensuremath{\mathrm{H}_2^+}}

\newcommand{\HDp}{\ensuremath{\mathrm{HD}^+\;}}
\newcommand{\HDpNoSpace}{\ensuremath{\mathrm{HD}^+}}

\newcommand{\Htwo}{\ensuremath{\mathrm{H}_2\;}}
\newcommand{\HtwoNoSpace}{\ensuremath{\mathrm{H}_2}}
\newcommand{\Hthreep}{\ensuremath{\mathrm{H}_3^+\;}}
\newcommand{\HthreepNoSpace}{\ensuremath{\mathrm{H}_3^+}}

\newcommand{\HtwopBe}{\ensuremath{\mathrm{H}_2^+ - {^9\mathrm{Be}}^+ \;}}
\newcommand{\HtwopBeNoSpace}{\ensuremath{\mathrm{H}_2^+ - {^9\mathrm{Be}}^+}}

\newcommand{\HthreepBe}{\ensuremath{\mathrm{H}_3^+ - {^9\mathrm{Be}}^+ \;}}
\newcommand{\HthreepBeNoSpace}{\ensuremath{\mathrm{H}_3^+ - {^9\mathrm{Be}}^+}}

\newcommand{\Ha}{\ensuremath{\mathrm{H}\;}}

\newcommand{\He}{\ensuremath{\mathrm{He}\;}}

\newcommand{\Hep}{\ensuremath{\mathrm{He}^+\;}}
\newcommand{\HepNoSpace}{\ensuremath{\mathrm{He}^+}}

\newcommand{\MHz}[1]{\SI{#1}{\mega\hertz}}

\newcommand{\um}[1]{\SI{#1}{\micro\meter}}
\newcommand{\nm}[1]{\SI{#1}{\nano\meter}}

\usepackage[normalem]{ulem}

\newcommand{\thetitle}{Trapping and Ground-State Cooling of a Single \Htwop}

\newcommand{\theauthors}{
\author{N. Schwegler}
\email{snick@phys.ethz.ch}
\author{D. Holzapfel}
\author{M. Stadler}
\author{A. Mitjans}
\author{I. Sergachev}
\author{J. P. Home}
\author{D. Kienzler}
 \email{daniel.kienzler@phys.ethz.ch}
\affiliation{
Institute for Quantum Electronics, Department of Physics, 
Eidgen\"ossische Technische Hochschule Z\"urich, \\
Otto-Stern-Weg 1, 8093 Zurich, Switzerland}
}

\begin{document}

\title{\thetitle}
\theauthors 

\begin{abstract}
We demonstrate co-trapping and sideband cooling of a \HtwopBe ion pair in a cryogenic Paul trap. We study the chemical lifetime of \Htwop and its dependence on the apparatus temperature, achieving lifetimes of up to $11^{+6}_{-3} \SI{}{\hour}$ at \SI{10}{\kelvin}. We demonstrate cooling of two of the modes of translational motion to an average phonon number of 0.07(1) and 0.05(1), corresponding to a temperature of \SI{22(1)}{\micro\kelvin} and \SI{55(3)}{\micro\kelvin} respectively. Our results provide a basis for quantum logic spectroscopy experiments of \HtwopNoSpace, as well as other light ions such as \HDpNoSpace, \HthreepNoSpace, and \HepNoSpace.
\end{abstract}

\maketitle

The hydrogen molecular ion is the simplest stable molecule and its internal structure can be calculated to very high precision, making it a valuable platform for determining fundamental constants and testing theory \cite{16Karr1,16Karr2}. 
However, \Htwop is difficult to study experimentally: its bound electronic excited states have negligible transition probabilities to low rovibrational states of the electronic ground state and, due to its non-polar nature, rovibrational transitions are dipole-forbidden resulting in extremely long lifetimes for most excited rovibrational states (weeks and longer) \cite{13Pilon,75Bishop}. These properties make direct laser cooling, internal readout by state-dependent fluorescence detection, and state-preparation by optical pumping impossible. Additionally, it readily reacts with \HtwoNoSpace, which is the primary background gas in ultra-high vacuum (UHV) systems, limiting trapping lifetimes. 

Only few contemporary \Htwop high-precision spectroscopy experiments exist, reflecting these experimental challenges. The LKB Paris experiment uses a Paul trap to confine $\approx 100$ \Htwop ions together with co-trapped \Be ions for sympathetic cooling with the goal of interrogating the fundamental vibrational transition. Rovibrational state-preparation is performed by resonance-enhanced multiphoton ionization (REMPI) of \Htwop and readout by state-dependent photo-dissociation \cite{2020Schmidt}. Trapping lifetimes of the \Htwop ions are limited by chemical reactions to a few minutes and their temperature is $\mathcal{O}(10) \, \SI{}{\milli\kelvin}$ \cite{PCHilico}. 
Similar techniques have enabled high-precision spectroscopy of the heteronuclear isotopologue \HDp and in combination with theory provided a determination of the proton-to-electron mass ratio to a relative uncertainty of $2\times10^{-11}$ \cite{2020Alighanbari,2020Patra}.

The ETH Z\"urich Molecular Physics and Spectroscopy group  utilizes Rydberg spectroscopy of a neutral \Htwo beam to extract properties of \Htwop via multichannel quantum-defect theory \cite{2018Beyer,2022Hoelsch}. This has enabled a measurement of the ortho-\Htwop hyperfine structure \cite{2004Osterwalder} and a determination of the first rotational interval of para-\Htwop with relative frequency uncertainty of $4.4\times10^{-7}$ \cite{15Haase}.

High-precision mass measurements have been performed using hydrogen molecular ions confined in Penning traps \cite{2015Myers,2017Hamzeloui,2020Rau,2020Fink,2021Fink}. In these experiments, the ions have been cooled to $\approx \SI{1}{\kelvin}$ by coupling to cryogenic electrical circuits and trapping lifetimes in excess of several months have been observed \cite{2021Fink}.

Optical spectroscopy of hydrogen molecular ions is expected to profit from experiments using trapped single \Htwop ions due to the suppression of systematic uncertainties. Selected rovibrational transitions in \Htwop are projected to reach a relative frequency uncertainty of $\mathcal{O}(10^{-17})$ \cite{14Schiller,16Karr1}. First steps towards high-precision spectroscopy of single hydrogen molecular ions have been demonstrated recently using co-trapped $\mathrm{HD}^+ - {^9\mathrm{Be}}^+$ pairs \cite{2021Wellers}. 
However single-ion experiments suffer from low signal and to be performed efficiently necessitate state preparation and a high spectroscopy duty cycle, i.e.\ reducing the duration or frequency of overhead operations like ion loading and state preparation. For \Htwop this challenge is amplified by several issues: It is created by ionizing \HtwoNoSpace, which requires some \Htwo density, but also reacts strongly with \Htwo to \HthreepNoSpace, limiting trapping lifetimes in case the ionization is performed in the trapping volume. 
Rovibrational state preparation will likely increase this challenge: Buffer-gas cooling with \He was proposed, but is slow and will thus likely require long \Htwop lifetimes \cite{17Schiller}.
Preparation of the \Htwop rovibrational state using REMPI typically requires an \Htwo beam intersecting with the trapping volume, further increasing the \Htwo concentration and thus lowering the \Htwop lifetime \cite{2020Schmidt}. These issues can be overcome by increased \Htwo pumping speed and long trapping lifetimes.

Quantum logic spectroscopy (QLS) has been shown to allow pure quantum-state preparation and quantum non-demolition readout even for hard to control ion species. In this technique a second, well-controlled ion species is co-trapped to exert control over the ion of interest through quantum gate operations which couple both ions to a shared normal mode of motion \cite{05Schmidt}. QLS has been used in the most accurate atomic clock and for quantum control and spectroscopy of highly-charged ions and molecular ions \cite{19Brewer,2020Micke,16Wolf,17Chou,2020Sinhal,2020Chou,2020Lin}.
It is a frequently suggested method to improve spectroscopy of \Htwop but has not been implemented thus far \cite{14Karr,16Karr1,16Leibfried,17Chou,17Schiller}.

In this letter we demonstrate trapping of single \HtwopBe ion pairs with a lifetime of $11^{+6}_{-3} \SI{}{\hour}$ and ground-state cooling of two of the normal modes of the ion pair's translational motion to a temperature of \SI{22(1)}{\micro\kelvin} and \SI{55(3)}{\micro\kelvin} respectively. 
Our achieved \Htwop lifetime should enable us to utilize buffer-gas cooling for rovibrational ground-state preparation and in combination with QLS to prepare pure quantum states of \Htwop as a starting point for high-precision spectroscopy \cite{17Schiller}. Ground-state cooling is a first step in many implementations of QLS and can be used to reduce the second-order Doppler shift caused by the ions' secular motion \cite{05Schmidt,2017Chen,2021King, wan2015efficient, 15Rugango}.

The ions are trapped in a monolithic, microfabricated linear Paul trap housed in a UHV chamber with an inner chamber cooled by a liquid helium flow cryostat. Single beryllium ions are loaded in the trap volume by photo-ionization of neutral beryllium atoms emitted from a thermal oven, while \Htwop is loaded through electron-impact ionization of residual gas molecules.
The choice of \Be as cooling ion is due to it being the lightest well-controlled ion species, providing sufficient participation of the two ions in shared normal modes of motion \cite{13Home}. 

Experiments primarily operate using the following sequence: Doppler cooling of \Be and initialization of ion order, \Be spin-state preparation by optical pumping, experiment-specific pulses, and finally \Be hyperfine state readout using state-dependent fluorescence captured by a photomultiplier tube \cite{98Wineland1}. Most operations on \Be use \nm{313} laser light to couple the \Be electronic ground state $S_{1/2}$ to the excited $P$ manifold. We use microwaves to drive magnetic dipole transitions in the \Be $S_{1/2}$ hyperfine manifold.

\begin{figure}[tb]
\includegraphics[width=0.48\textwidth]{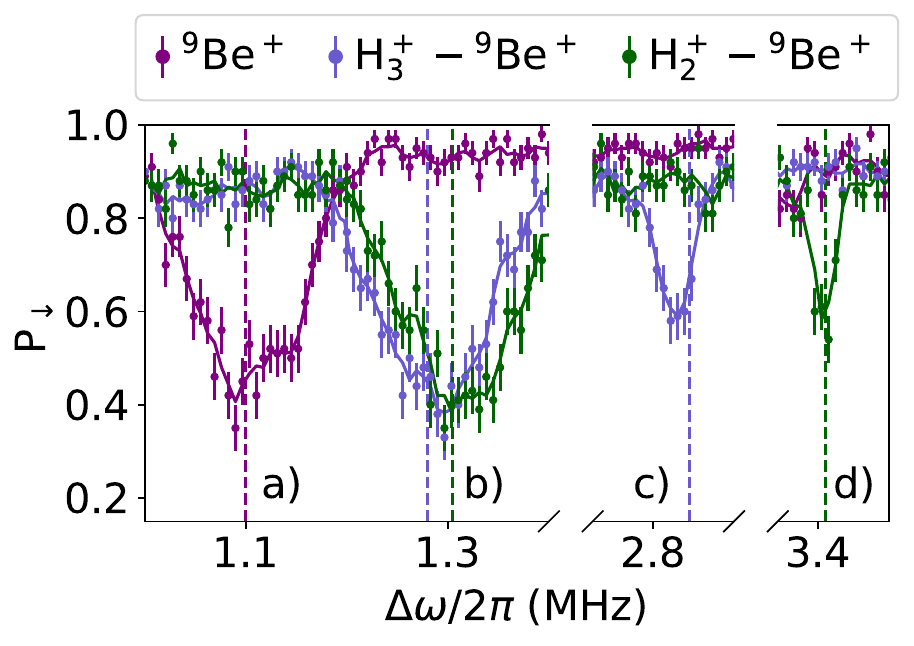}
\caption{\label{fig:H2_lifetime_example} Example of sideband spectroscopy scans on \Be probing the axial mode frequencies to distinguish \HtwopBeNoSpace, \HthreepBeNoSpace, and \BeNoSpace. The detuning of the Raman lasers from the carrier transition is denoted $\Delta \omega$ and the measured \Be \ket{\downarrow} probability P$_{\downarrow}$. The solid lines represent a running average over three points to guide the eye. The vertical dashed lines mark the calculated frequency of the in- and out-of-phase axial mode of \HtwopBe and \HthreepBe based on the axial frequency of a single \Be ($2\pi\times\SI{1.1}{\mega\hertz}$) confined in the same trapping potential. Different frequency ranges use different probe duration to optimize signal strength. The sets for \HtwopBe and \HthreepBe are taken a few minutes apart, during which \Htwop underwent a chemical reaction to \HthreepNoSpace. The mass changed from \SI{2} to \SI{3}{\atomicmassunit}, which shifts b) the axial in-phase mode at $2\pi\times\SI{1.3}{\mega\hertz}$ by $-2\pi\times\SI{25}{\kilo\hertz}$, and the axial out-of-phase mode is shifted from d) $2\pi\times\SI{3.4}{\mega\hertz}$ to c) $2\pi\times\SI{2.8}{\mega\hertz}$. A single \Be shows a resonance in scan a) only, giving us a means of detecting loss of \HtwopNoSpace. The duration to complete one set of data containing all four scans is $\SI{80}{\second}$. 
}
\end{figure}

We implement loading of \HtwopBe by first trapping a single \Be ion, which we detect by imaging its fluorescence light onto a CMOS camera. By performing real-time image analysis, we continuously monitor its axial position. To load an \Htwop ion we use an electron beam which ionizes the residual gas in the vacuum chamber. When an ion is loaded, it is sympathetically cooled by the \Be ion in the trap. We cannot observe this `dark ion' directly as it does not fluoresce, but since sympathetic cooling causes the ion to crystallise, we detect a shift in the \Be position on the camera image. Since electron bombardment is not species-selective we observe loading of several species: We dominantly load \HepNoSpace, but also observe loading of ions heavier than \BeNoSpace, compatible with a mass range of $\approx 14-\SI{20}{\atomicmassunit}$ (assuming singly-charged cations). We mass-selectively remove parasitic ion species by excitation of their motional modes.

More details on the apparatus, control of \HtwopBeNoSpace, and loading procedure can be found in the supplemental material, which includes Refs.~\cite{2019Ragg, 2014Lo, 2015Guggemos, 2019Meir}.

To identify the presence of \HtwopNoSpace, we perform motional sideband spectroscopy on the \Be spin using a pair of Raman beams. We probe the blue motional sideband spectrum and, in case \Htwop was loaded, can identify the expected resonances of the axial in- and out-of-phase modes of \HtwopBeNoSpace. We observe drifts in the sideband resonances of up to $\approx 2\pi \times \SI{10}{\hertz\per\s}$, which are likely caused by changing electrical stray fields originating from surfaces charged during the loading procedure. Comparing the in- and out-of-phase resonance frequencies to normal mode calculations enables us to determine both charge and mass of the second trapped ion without precise knowledge of the strength of the trapping potential. The calculations only assume trapping of a single \Be ion and a second ion of variable charge and mass. Using this method, we can distinguish trapping of \HtwopNoSpace, \HthreepNoSpace, and \HepNoSpace, and also exclude potential trapping of e.g.\ $\mathrm{He}^{2+}$. Example sideband spectra for \BeNoSpace, \HtwopBeNoSpace, and \HthreepBe are shown in Fig.~\ref{fig:H2_lifetime_example}.

\begin{figure}[tb]
\includegraphics[width=0.48\textwidth]{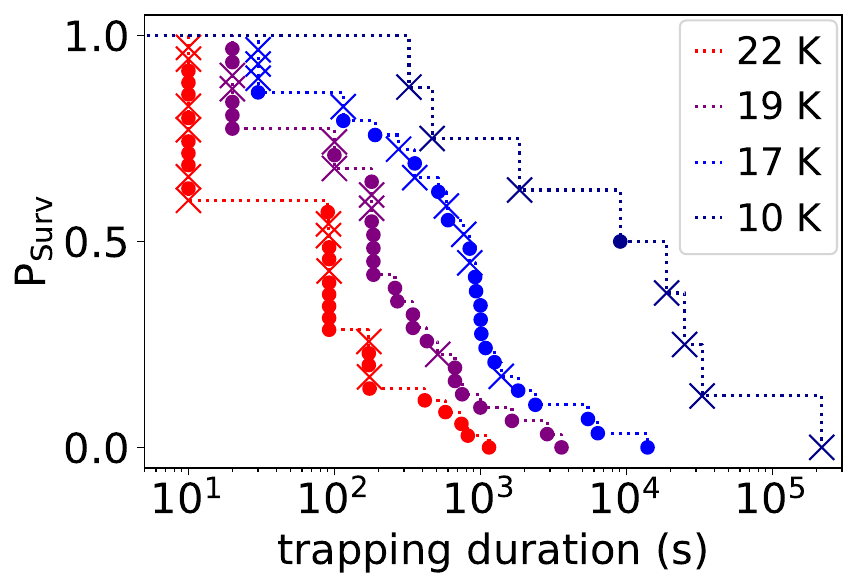}
\caption{\label{fig:H2_lifetime_data} Survival probability $P_{\text{Surv}}$ of \HtwopBe measured for different cryostat temperatures. For clarity, the data at \SI{22}{\kelvin} (\SI{19}{\kelvin}, \SI{17}{\kelvin}) is offset by $\SI{10}{\second}$ ($\SI{20}{\second}$, $\SI{30}{\second}$). Chemical reactions $\Htwop + \Htwo \rightarrow \Hthreep + \mathrm{H}$ are marked with `$\bullet$', while ion loss is marked with 
`$\times$'. The data is not pre-processed and no additional loss mechanisms are observed.}
\end{figure}

Due to the limited \Htwo pumping speed in UHV chambers, the reaction $\Htwop + \Htwo \rightarrow \Hthreep + \mathrm{H}$ is usually the dominant loss mechanism for \Htwop and limits the trapping lifetime at typical UHV pressures to a few minutes \cite{PCHilico}. The cooled inner chamber allows us to reduce the \Htwo partial pressure by cryogenic pumping, extending \Htwop trapping lifetimes.

\begin{figure}[tb]
\includegraphics[width=0.48\textwidth]{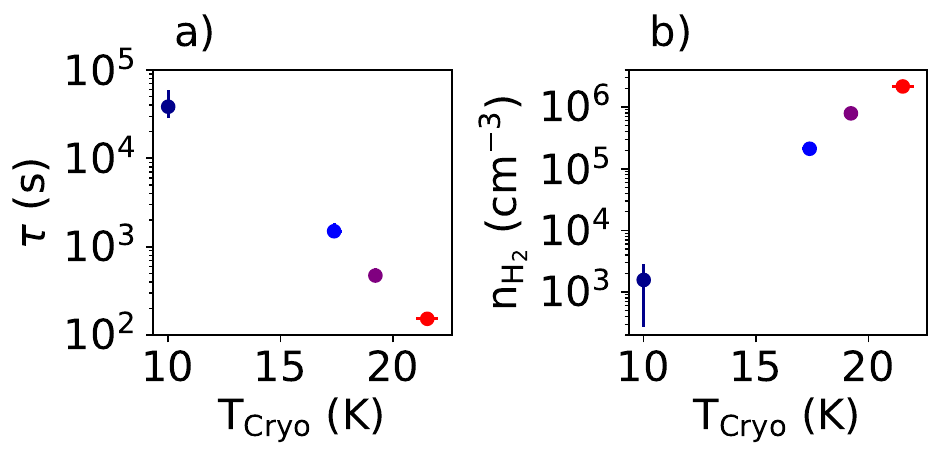}
\caption{\label{fig:H2_lifetime_average_pressure} a) Lifetime $\tau$ of \HtwopBe and b) partial \Htwo density $n_{\mathrm{H}_2}$ for different cryostat temperatures, extracted from the data shown in Fig.~\ref{fig:H2_lifetime_data}. The error bars on the y-axis show the 68 \% confidence interval and are derived assuming an exponential distribution. The error bars on the temperature values indicate the standard deviation of the measured temperatures and are limited by the precision of the temperature control loop.}
\end{figure}

To investigate the lifetime of \Htwop in our apparatus, we load an \HtwopBe pair and continuously perform measurements to identify the trapped ions. We observe three loss mechanisms: Chemical reaction $\Htwop + \Htwo \rightarrow \Hthreep + \Ha$, loss of \HtwopNoSpace, and loss of \BeNoSpace. To distinguish these cases, we use the \Be fluorescence signal to confirm the presence of \Be and again perform motional sideband spectroscopy on \Be to discriminate single \BeNoSpace,  \HtwopBeNoSpace, and \HthreepBeNoSpace. Since all signals rely on \BeNoSpace, we cannot measure if \Htwop is still trapped after \Be was lost. 
We define the trapping duration of the \HtwopBe pair as the time difference between the first and last scan showing the signature of \HtwopBeNoSpace.

We repeat the ion-loss measurement several times and calculate the survival probability $P_{\text{Surv}}(t)$ of the \HtwopBe pair after a trapping duration of $t$ as $P_{\text{Surv}}(t) = N(t) / N_{\text{tot}}$, where $N(t)$ is the number of ion-loss measurements with trapping duration larger or equal to $t$ and $N_{\text{tot}}$ is the total number of ion-loss measurements. We took four data sets at cryostat temperatures of \SI{10}, \SI{17}, \SI{19}, and \SI{22}{\kelvin}, which are shown in Fig.~\ref{fig:H2_lifetime_data}.
Estimations of the thermal conductance and heat load in our apparatus suggest that the corresponding temperature of the inner chamber housing the trap is $< \SI{5}{\kelvin}$ warmer than the measured cryostat temperature. At lower temperatures we observe the expected reduced chemical reaction rate, which is consistent with increased cryogenic pumping of the residual \Htwo gas. While ion loss is present at all temperatures and also reduces with lower cryostat temperature, it is the dominant loss channel for the \SI{10}{\kelvin} data set. The entire data set (all cryostat temperatures combined) contains a total of 103 events, 68 of which are chemical reactions, 24 are loss of \Htwop and 11 are loss of \BeNoSpace. Of the \Be loss events, seven can be directly correlated with issues of the experiment control system and stability of the \Be cooling laser. The mechanism responsible for \Htwop loss remains unclear. In separate experiments where we block the \Be cooling laser for a certain duration, we observe \Htwop loss while \Be stays trapped. The uncooled trapping duration of \Htwop was a few seconds during the time the data set was acquired. We have observed an improvement to approximately \SI{10}{\second} since then.

From the survival probability data we derive the \HtwopBe lifetime, shown in Fig.~\ref{fig:H2_lifetime_average_pressure} a). We assume an exponential distribution to simplify the estimation of the error bars, albeit the observed survival probabilities seem to follow a more complex distribution. The lifetime, which includes all loss channels, increases from only a few minutes at \SI{22}{\kelvin} to $\tau = 11^{+6}_{-3} \SI{}{\hour}$ at \SI{10}{\kelvin}. 
Because we can distinguish ion loss from the $\Htwop + \Htwo \rightarrow \Hthreep + \Ha$ reaction we can calculate the reaction rate for each cryostat temperature by dividing the total observation time by the number of observed reactions. The reaction rate coefficient is known and well approximated by the Langevin collision rate coefficient $k_{L} = \SI{2.1e-9}{\centi\meter\cubed\per\second}$ \cite{asvany2009numerical, glosik1994measurement}. This allows us to estimate the partial \Htwo density $n_{\mathrm{H}_2}$ in the trap volume in dependence of the cryostat temperature, shown in Fig.~\ref{fig:H2_lifetime_average_pressure} b). At a cryostat temperature of \SI{10}{\kelvin} we observe a density of $n_{\mathrm{H}_2} = 1.6 (1.3) \times 10^{3}$  \SI{}{\per\cubic\centi\metre}. Assuming a temperature of the \Htwo gas of \SI{15}{\kelvin} for this density results in an \Htwo partial pressure of $p_{\mathrm{H}_2} = 3.3(2.7)\times 10^{-13}$ \SI{}{\pascal}.

Ground-state cooling of the translational motion of two-ion chains containing a spectroscopy and readout ion is an often-used ingredient for implementing QLS and can also be used to reduce the second-order Doppler shift in spectroscopy caused by the ion's secular motion \cite{05Schmidt}. Due to the large mass mismatch between \Htwop and \BeNoSpace, each radial mode has a strong amplitude for only one of the two species, whereas the other ion only participates weakly. In contrast, the two axial modes exhibit a stronger participation of both ions simultaneously, which makes them suitable candidates for the transfer mode in quantum logic protocols \cite{2021King}. 
The out-of-phase mode can have advantages for implementing QLS, such as typically reduced heating rates and different mode participation of the ions compared to the in-phase mode. However, in our system, a large \Be Lamb-Dicke parameter of the in-phase mode causes poor contrast on operations on the out-of-phase mode due to the Debye-Waller effect \cite{98Wineland1}. To achieve high fidelity in QLS operations using the out-of-phase mode, the in-phase mode has to be ground-state cooled as well. We demonstrate ground-state cooling of the axial in-phase mode ($2 \pi \times \SI{1.3}{MHz}$) and axial out-of-phase mode ($2 \pi \times \SI{3.4}{MHz}$) of \HtwopBe using resolved-sideband cooling on \Be \cite{89Diedrich}. Because of the high Lamb-Dicke parameter for our Raman beam geometry of $\eta = 0.57$ for the in-phase mode, the thermal state occupation after Doppler cooling is not in the Lamb-Dicke regime and direct cooling to the ground state using the first red sideband only is not possible. Thus, we first use the fourth red sideband to reduce the population to below $n=4$, followed by sideband cooling using the first red sideband. We implement this in a pulsed fashion, alternating red-sideband pulses with microwave and repumping pulses to reset the spin to \ket{\downarrow}. 
The out-of-phase mode with a Lamb-Dicke parameter of $\eta = 0.1$ can be cooled with only the first red sideband. For simultaneous ground-state preparation of the two axial modes, we first perform ground-state cooling of the in-phase mode as described above, and subsequently ground-state cool the out-of-phase mode. During this step, the in-phase mode gets excited by recoil from the repumping pulses. We therefore follow the out-of-phase cooling with a repetition of the in-phase first-sideband cooling sequence.
We estimate the average motional population $\nbar$ after both Doppler and ground-state cooling by performing red- and blue-sideband spectroscopy and using the sideband ratio method \cite{98Wineland1}, extracting the sideband contrast through fits to the data and correcting for \Be state-preparation and readout error using data from reference Rabi oscillations driven by microwaves. For the in-phase mode we obtain $\nbar = \num{10(5)}$ after Doppler and $\nbar = \num{0.07(1)}$ after ground-state cooling. For the out-of-phase mode we obtain $\nbar = \num{2.4(5)}$ after Doppler and $\nbar = \num{0.05(1)}$ after ground-state cooling. The sideband- and Doppler-cooled data for both the in- and out-of-phase mode is shown in Fig.~\ref{fig:sbc}. Since the data was taken at the $\pi$-time of the corresponding sideband in the motional ground state, it demonstrates the achievable signal contrast for readout of the transfer mode in quantum logic protocols. 

\begin{figure}[tb]
\centering
\includegraphics[width=0.48\textwidth]{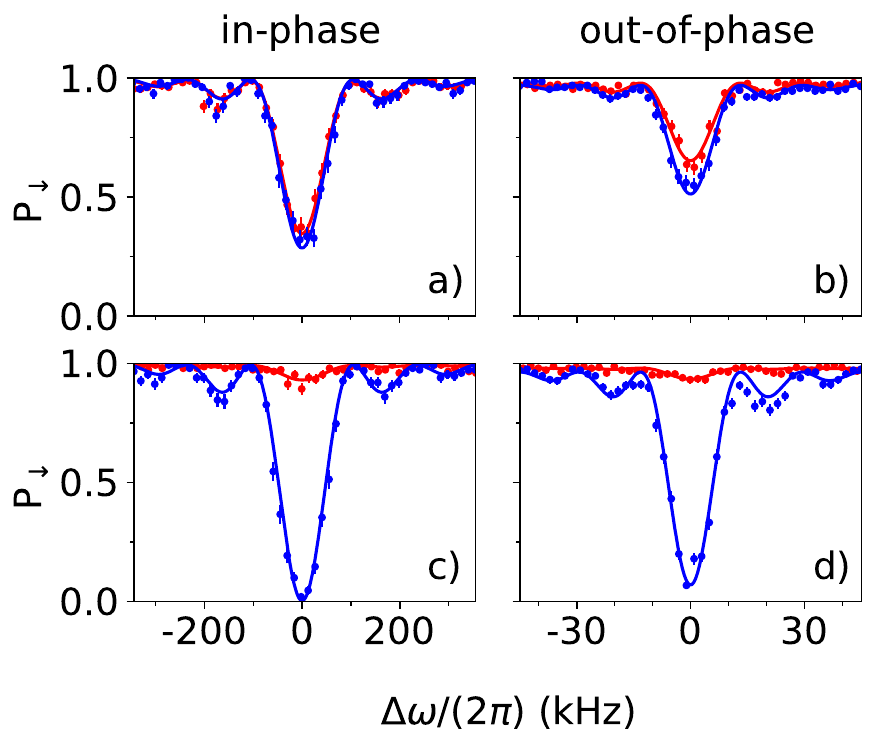}
\caption{\label{fig:sbc} Red and blue sidebands of the axial in- and out-of-phase modes of motion of \HtwopBe probed using the stimulated Raman transition on \Be. The detuning of the Raman lasers from the corresponding sideband transition is denoted $\Delta \omega$ and the measured \Be \ket{\downarrow} probability $P_{\downarrow}$. The fits (solid lines) are used to extract the contrast and determine the average phonon number. We obtain for the in-phase mode a) $\nbar = \num{10(5)}$ after Doppler cooling and c) $\nbar = \num{0.07(1)}$ after sideband cooling. For the out-of-phase mode we obtain b) $\nbar = \num{2.4(5)}$ after Doppler cooling and d) $\nbar = \num{0.05(1)}$ after sideband cooling. For the data set c) the out-of-phase mode is Doppler cooled, while for the data set d) both the in- and out-of-phase mode are cooled to the ground state simultaneously. In the latter case, the temperature of the in-phase mode is similar to the one obtained from c).}
\end{figure}

By performing the measurement of $\nbar$ with different wait times after ground-state cooling, we derive a heating rate from the ground state of $\SI{9.3(4)}{quanta \per s}$ for the axial in-phase mode and $\SI{10.6(7)}{quanta \per s}$ for the axial out-of-phase mode. For the axial motion of a single \Be ion (mode frequency $2 \pi \times \SI{1.1}{MHz}$) we measure a heating rate of $\SI{4.2(3)}{quanta \per s}$. These heating rates are low enough such that the axial in- and out-of-phase modes of \HtwopBe can be used as transfer modes in quantum logic protocols, since they do not heat up considerably for an expected \Htwop state readout duration \cite{17Chou}.

We have reported on trapping and ground-state cooling of a \HtwopBe ion pair, paving the way towards quantum control and high-precision spectroscopy of \Htwop using QLS. Future implementation of QLS will require rovibrational state-preparation of the \HtwopNoSpace. The electron-impact ionization that was used for this study creates the \Htwop in a distribution of rovibrational states \cite{93Weijun}. With \He buffer-gas cooling, it should be possible to prepare the rovibrational ground-state of \Htwop \cite{17Schiller}, and our apparatus has the capabilities necessary to test this approach. Another possibility may be to use REMPI to create \Htwop in the desired rovibrational state \cite{2020Schmidt, zhang2023generation}. Beyond \HtwopNoSpace, our apparatus and techniques are also suitable to control other light ion species, such as \HepNoSpace, \HthreepNoSpace, and \HDpNoSpace. 

The authors thank B. MacDonald-de Neeve and C. Axline for technical support, the Segtrap and Penning teams of the ETHZ TIQI group for sharing laser light, and D. Leibfried, C.-W. Chou, D. R. Leibrandt, D. B. Hume, Ch. Kurz, D. J. Wineland, D. T. C. Allcock, D. H. Slichter, S. Willitsch, L. Hilico, J.-Ph. Karr, F. Merkt and M. Grau for helpful discussions and advice. This work was supported by Swiss National Science Foundation Grant No. 179909, by ETH Research Grant No. ETH-52 19-2, as a part of NCCR QSIT, a National Centre of Competence (or Excellence) in Research, funded by the Swiss National Science Foundation (grant number 51NF40-185902), and by EU Quantum Flagship H2020 FETFLAG-2018-03 under Grant Agreement No. 820495 AQTION. 

\bibliographystyle{apsrev4-2}
\bibliography{refs}

\newpage
\clearpage
\pagebreak
\bibliographystyle{apsrev4-2}
\renewcommand{\bibliography}[1]{}

\title{Supplemental material for:\\\thetitle}
\theauthors 
\maketitle

\section{Apparatus}
\label{appendix_apparatus}
The trap is fabricated by laser-enhanced etching of a fused silica substrate. The electrodes are defined by electron-beam physical vapor deposition and electromechanical deposition of gold~\cite{2019Ragg}. The trap is monolithic with an electrode-ion distance of \um{300}. We operate the trap with a dual-phase RF drive at a frequency of $\Omega_{\mathrm{rf}} = 2 \pi \times \MHz{64.3}$ The RF drive is realized with a resonant circuit built from discrete components on a printed circuit board. \bC{The frequencies and amplitudes of the motional modes of the \HtwopBe ion pair corresponding to our experimental trapping conditions are shown in Table \ref{BeH2crystal}}.

\begin{table}[h]
\centering
\caption{\bC{Calculated frequencies and amplitudes for the radial (x,y) and axial (z) motional modes for an \HtwopBe ion pair corresponding to our experimental trapping conditions. The respective in-/out-of-phase modes are denoted as `IP'/`OP'. The \HtwopBe pair aligns along the axial direction.}}
\label{BeH2crystal}
\begin{tabular}{ccccc}
\hline \hline
 &  & $\omega/2\pi$ & \Be & \Htwop \\
Axis & Mode & (\SI{}{\mega\hertz}) & Amplitude (\SI{}{\nano\meter}) & Amplitude (\SI{}{\nano\meter}) \\
\hline
x & IP & 9.46 & 0.1 & 16.3 \\
y & IP & 9.32 & 0.1 & 16.5 \\
x & OP & 1.92 & 17.1 & 0.5 \\
y & OP & 1.76 & 17.9 & 0.6 \\
z & OP & 3.41 & 3.5 & 26.2 \\ 
z & IP & 1.30 & 20.0 & 11.8 \\ 
\hline \hline
\end{tabular}
\end{table}

The chamber consists of two parts: An outer vacuum chamber, and an inner chamber containing the trap, thermally connected to the coldhead of a flow cryostat (Janis ST-400). We built the chamber following standard UHV procedures, using exclusively low-outgassing material (with the exception of a charcoal getter for improved cryogenic pumping), and baking at $\approx \SI{110}{\degreeCelsius}$ for several days. The vacuum chamber features a combination of ion and a non-evaporable getter pump (SAES Group NexTorr D500 Starcell) and a leak valve for the introduction of He gas for buffer gas cooling. As discussed in Sec.~\ref{appendix_loading}, there are indications that the partial He pressure is elevated in the chamber. We suspect that this is caused by residual outgassing of He trapped in the charcoal getter or other parts of the system after testing the leak valve. The flow cryostat has a liquid helium consumption of $\approx \SI{0.9}{\liter\per\hour}$ when operated at \SI{10}{\kelvin}. The cryostat temperature is measured in the coldhead and depends on the flow rate of liquid helium. We regulate the coldhead temperature by changing the flow with a stepper motor that operates a needle valve on a transfer line supplying liquid helium to the cryostat from a storage dewar. The line of sight of the central trap region to the \SI{300}{\kelvin} part of the chamber was minimized to reduce background gas pressure at the trapped ions' location. Openings in the inner chamber with line-of-sight are for an electron and a beryllium beam, emitted from sources mounted in the \SI{300}{\kelvin} part. The combined solid angle of the openings is \SI{0.06}{\steradian}. Further openings without line of sight provide (limited) vacuum pumping conductivity and a path for leaked in He gas to enter the trapping region. Optical access for laser beams and fluorescence imaging is provided by openings covered with glass windows (imaging) and lenses (beam access).

\section{Control of \HtwopBeNoSpace}
\label{appendix_Be_control}
\bC{We load single} beryllium ions 
\bC{in the trap volume} 
by photo-ionization of neutral beryllium atoms evaporated from an electrically heated oven 
\bC{using a continuous-wave \SI{235}{\nano\meter} laser} \cite{2014Lo}.

\begin{figure}
\includegraphics[width=0.48\textwidth]{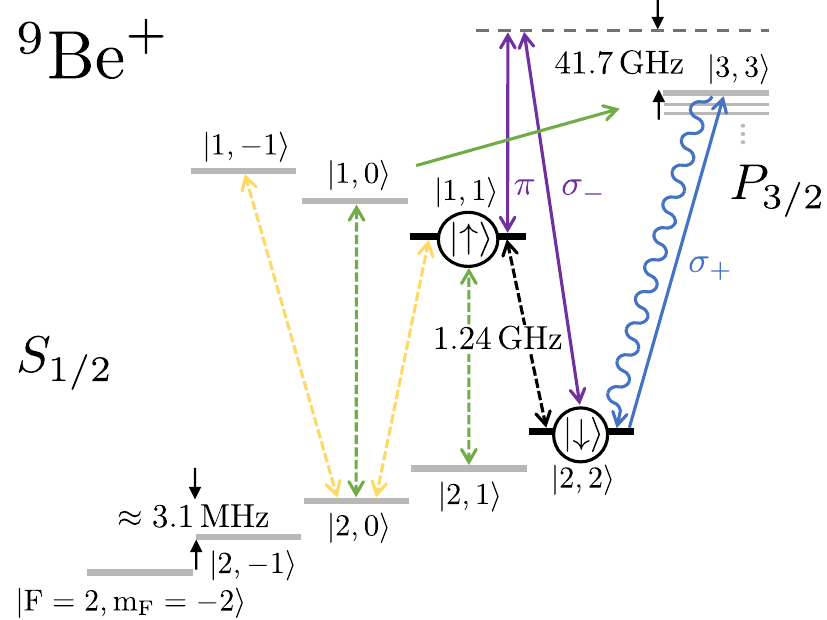}
\caption{\label{fig:Be_level_diag} \bC{\Be level diagram with microwave (dashed arrows) and laser transitions (solid arrows). The closed cycling transition is shown in blue and is used for Doppler cooling and spin-dependent fluorescence readout. Arrows in green show the optical pumping scheme used for electronic state preparation in $\ket{\downarrow}$. Arrows in yellow indicate the microwave transitions used for shelving prior to readout. The Raman transition for sideband spectroscopy and ground-state cooling is shown in purple. The splitting between the Zeeman-sublevels in the S$_{1/2}$ manifold ranges from \SIrange[range-units = single]{3.13}{3.17}{\mega\hertz} for our magnetic field of \SI{0.45}{\milli\tesla}.}}
\end{figure}

For cooling, state preparation, coherent control and readout of \Be we use a combination of microwaves and continuous-wave laser at \SI{313}{\nano\meter} \bC{as shown in Fig.~\ref{fig:Be_level_diag}.} The microwave transitions are driven from an oscillating magnetic near-field created by a current in a conductive line located $\approx \SI{5}{\milli\meter}$ away from the trap center. We use the two-level system defined by the level pair $\ket{\downarrow} \equiv \ket{S_{1/2}, F=2,m_{F}=+2}$ and $\ket{\uparrow} \equiv \ket{S_{1/2}, F=1,m_{F}=+1}$, with a splitting of $2 \pi \times \SI{1.24}{\giga\hertz}$. The quantization axis is provided by a magnetic field of \SI{0.45}{\milli\tesla}. 

The following sequence is used in most experiments: Doppler cooling of \Be and initialization of ion order, \Be state preparation to \ket{\downarrow}, experiment-specific pulses, \Be fluorescence readout \bC{using a photomultiplier tube}. In the following, each of the steps is explained in more detail.

Using a laser resonant with the closed-cycling transition $\ket{S_{1/2}, F=2,m_{F}=+2} \leftrightarrow \ket{P_{3/2}, F=3,m_{F}=+3}$ (pointing along the quantization axis with $\sigma_{+}$ polarization at approximately half a saturation intensity) we implement spin-state-dependent fluorescence readout of \BeNoSpace, where $\ket{\downarrow}$ is bright and $\ket{\uparrow}$ is dark. Prior to the readout, the population of the dark state $\ket{\uparrow}$ is transferred to the $\ket{S_{1/2}, F=1,m_{F}=\bC{-}1}$ level using two microwave pulses to minimize off-resonant excitation.

The motion of the \Be ion is Doppler cooled in two steps. All laser beams used for Doppler cooling point along the quantization axis, are $\sigma_{+}$ polarized and have a projection along all normal modes of the trapped ions. In the first step a higher power, far-detuned `pre-cooling' laser beam provides optical pumping into $\ket{\downarrow}$ and robust cooling to efficiently (re-)crystallize hot ions during loading or after background gas collisions ($\approx \SI{600}{\mega \hertz}$ detuned from the cycling transition, $\approx \SI{0.75}{\milli\watt}$ power and a beam waist radius of $\approx \SI{10}{\micro\meter}$, corresponding to $\sim 1000$ saturation intensities). 
In the second step a laser beam with lower power and closer to resonance (detuned by approximately half the linewidth at approximately half the saturation intensity) further cools the ions motion close to the Doppler limit. To prevent population leakage to the $S_{1/2}, F=1$ manifold, 
we simultaneously turn on a `repumping' beam, which couples the $S_{1/2}, F=1$ manifold to the $P_{3/2}$ manifold. This beam has an intensity much lower than one saturation intensity.

The electronic state of \Be is initialized into $\ket{\downarrow}$ using a combination of microwave pulses and optical pumping by the repuming beam.

Stimulated Raman transitions driven by two laser beams blue detuned by $2\pi\times\SI{41.7}{\giga\hertz}$ from the transition $\ket{S_{1/2}, F=2,m_{F}=+2} \leftrightarrow \ket{P_{3/2}, F=3,m_{F}=+3}$ implement coherent spin-motion interaction for motional sideband spectroscopy and ground-state cooling (Carrier Rabi oscillations with pi time of $\approx \SI{4.5}{\micro\second}$). One of the beams points along the quantization axis and is $\sigma_{\bC{-}}$ polarized, while the other is perpendicular and $\pi$ polarized. This geometry creates coupling to only the axial modes of motion. 

Occasional collisions with background gas can swap the positions of the two ions in a \HtwopBe pair. 
To ensure the same ion order for each experiment, the order is initialized during the pre-cooling pulse by radially displacing the ions with an electric field created by a voltage bias applied to the trap RF electrodes. \bC{Due to its higher mass, the \Be is displaced further, rotating the crystal such that it is approximately radially oriented. Upon removal of the voltage bias, we observe a particular ion order, which means that there likely is a tilt of the DC confinement with respect to the RF confinement present in the trap.}~\cite{13Home}.

\section{\HtwopBe loading procedure}
\label{appendix_loading}
\Htwop is produced by electron-impact ionization of \Htwo from rest gas. A commercial cathode (Kimball ES-042 Tantalum Disc Cathode) combined with custom parts for electron extraction and beam collimation creates an electron beam with an energy of \SI{285}{\electronvolt} and an emission current of \SIrange[range-units = single]{10}{40}{\micro\ampere}. To switch the electron-impact ionization on and off, we control the current in the magnetic field coils that provide the quantization field. When the coils are turned on, we do not observe ionization of background gas, indicating that most of the electrons are deflected away from the trap. During the electron-bombardment, we turn on the \Be pre-cooling laser (see Sec.~\ref{appendix_Be_control} for details) to sympathetically cool hot dark ions. When a dark ion crystallizes, it shifts the \Be ion position by $\approx \SI{4.5}{\micro\meter}$, which is recognized by a script that traces the ion's position in real-time from camera images. Several experiments that co-trap a single cooling ion and a dark ion have observed  crystallization times after ionization on the order of minutes and longer \cite{2015Guggemos,2019Meir}. We have not observed crystallization of dark ions after the electron beam was turned off, which, on the contrary, suggests that the time between ionization and crystallization in our apparatus is on the order of seconds.

During \Htwop loading, the electron bombardment also leads to ionization and loading of other ion species from background gas. The motional mode spectrum of the mixed species ion pair is dependent on the mass and charge of the trapped ions, which allows to distinguish between different ion pairs by probing the axial motional sidebands, see main text for an example. 
We have observed co-trapping of dark ions with mass \SI{2}{\atomicmassunit} (\HtwopNoSpace), \SI{3}{\atomicmassunit} (\HthreepNoSpace), \SI{4}{\atomicmassunit} (\HepNoSpace) and heavier ions $\approx 14-\SI{20}{\atomicmassunit}$ (assuming singly-charged cations). The loading is dominated by \HepNoSpace, indicating that the He partial pressure is elevated following a test of the leak valve, see Sec.~\ref{appendix_apparatus}. 

To ensure the loading of a \HtwopBe pair, we mass-selectively remove the parasitic ion species by excitation of motional modes. We do this by applying an additional RF signal to the RF electrodes, producing an oscillating electric field at the ion that coherently excites the motional mode on resonance. We refer to this technique as 'tickle' in the following.

To prevent \Hep from being trapped, we continuously repeat a tickle scan during the electron-bombardment to excite one of the \Hep radial motional modes at $\approx 2\pi\times\SI{4.7}{\mega\hertz}$. In each scan, we use a total of 20 tickle pulses, each with a duration of $\SI{100}{\micro\second}$ and a different frequency offset from the resonance. For the first 10 pulses of the tickle scan, the frequency offset is varied from 0 to $+2\pi\times\SI{50}{\kilo\hertz}$. For the next 10 pulses, the frequency offset is varied from 0 to $-2\pi\times\SI{50}{\kilo\hertz}$. 
We remove dark ions with a mass larger than \Be after crystallization. We can detect that a co-trapped dark ion is heavier than \Be by reordering the ion crystal, see Sec.~\ref{appendix_Be_control}. When a heavier ion is co-trapped, the \Be initializes in the opposite position compared to the case where a lighter ion is co-trapped.
To remove the co-trapped heavy dark ion, we use the experiment sequence described in Sec.~\ref{appendix_Be_control} with a fixed-frequency tickle pulse at $2\pi\times\SI{0.85}{\mega\hertz}$, a frequency we believe is resonant with the heavy ion's radial trap frequency. To stop the \Be from sympathetically cooling the dark ion, we block the \Be cooling beams during the experiments for a few seconds, and then unblock it to detect whether the dark ion is still trapped. We observe that within a few repetitions the dark ions leaves the trap.

The loading sequence for a \HtwopBe pair is summarized in the following:
\begin{enumerate}
    \item Load a \bC{single} \Be ion. 
    \item Turn on continuous far-detuned \BeNoSpace cooling beam and start \HepNoSpace~tickle sequence. Turn on emission of electron gun. 
    \item Switch magnetic field to steer electrons into the trapping region. 
    \item Through electron-impact ionization, a dark ion is produced from background gas and sympathetically cooled by the \BeNoSpace. Upon crystallization, the dark ion is detected. Switch the magnetic field to deflect the electrons to prevent further ionization of the rest gas. Turn off the \HepNoSpace tickle. 
    \item Perform reorder to probe whether the captured dark ion is lighter or heavier than \BeNoSpace.
    \begin{enumerate}
        \item If a heavier dark ion was loaded, remove it from the trap by tickling. Restart \HepNoSpace tickle and return to point 3 to load another dark ion. 
        \item If a lighter dark ion was loaded, turn off emission of the electron gun. Perform a set of sideband spectroscopy scans to probe the axial trap frequencies and verify the correct mass ratio associated with \HtwopBeNoSpace.
    \end{enumerate}
\end{enumerate}

The rate of ionization is proportional to the density of background gas and emission current of the electron gun. Depending on the cryostat temperature, we observe different loading rates for the same emission current, indicating different partial pressures in the trapping region, as verified for \Htwo in the main text. For a given temperature, we select the highest emission current such that we still operate in a regime where a \HepNoSpace can be removed by motional excitation before the next dark ion is trapped. We observe that the \Be ion is lost approximately once during the loading process and has to be reloaded. Our loading rate of \Htwop is currently limited by the overhead from reloading \Be and the ratio between the background gas density of \Htwo and He. Loading a \HtwopBe pair typically takes $\approx \SI{15}{\minute}$ at a cryostat temperature of \SI{10}{\kelvin}. 

\bibliographystyle{apsrev4-2}
\bibliography{refs}

\end{document}